# Strength and structure of carbon–carbon reinforced composite

Vitaly Chaban[1]

Department of Chemistry, University of Rochester, Rochester, NY 14627, USA

**Abstract.** The atomistic simulations of carbon nanotube (CNT) – carbon reinforced composite material are reported. The studied composite samples were obtained by impregnating certain amounts of CNTs (3,3) and (6,6) into pristine graphite matrix. The addition of CNTs is found to be of significant usefulness for nanotube–reinforced carbon composites since it allows to achieve extreme lightness and strength. Being impregnated into graphite matrix, nanotubes are able to increase the critical component of its highly anisotropic Young modulus by 2-8 times. The linear thermal expansion coefficients do not exceed $10^{-6}$-$10^{-5}$ $K^{-1}$, making these composites applicable for aviation and space vehicles. The dispersion of CNTs within graphite matrix was found to drastically influence composite properties.

[1]Corresponding author. E-mail: v.chaban@rochester.edu

**Introduction**

An intriguing ability of graphite and carbon nanotubes (CNTs) to be impregnated with different nanosized structures [1-5] as well as their extreme strength [6,7] give rise to a great variety of new composite materials [8-10]. Extremely high porosity resulting in low density, large surface area and multiple surface defects are all among the factors making graphite an indispensable candidate for composite applications. Moreover, expanded graphite, with pores ranging from 1.5 nm to a few μm, is known to be compacted without a binder over a wide range of densities. In turn, recent applications of CNTs as reinforcing agents of the composites are widespread [11-13]. Note, many parameters vigorously influence the effective properties of the nanotube reinforced composites, e.g. CNT particular structure, orientation within composite matrix, dispersion degree, diameter and length distributions, overall matrix stiffness, etc. Qian et al. [14] reported a 35 to 42 % increase in the elastic modulus and a 25 % increase in strength by adding only 1 wt. % CNTs to the conventional polyester resin. Meanwhile, numerous experimental and theoretical studies have been performed [15-27] to determine the mechanical properties of the CNTs since their discovery, although providing contradicting outcomes. The comprehensive review of the mechanical properties of carbon nanotubes was recently published by Shokrieh and Rafiee [8]. Significant scattering in the published results can be attributed to the differences in investigation techniques as well as to prerequisites of calculations. Computer simulations studies [23-24] usually deal either with an isolated nanotube or a nanotube in polymer matrix considering predefined mutual orientation of both. Meanwhile, by means of modern experimental techniques macroscopic samples are investigated [25-27].

The idea of manufacturing a composite material containing graphite as a matrix and generally carbon fibres as a reinforcing agent is rather old, and first appeared in the second part of the last century. Such materials are usually referred to as carbon fibre – reinforced carbon (RCC) composites. Unlike other composites, both the reinforcement and matrix phases consist of essentially pure carbon. The list of desirable properties of RCC include an extremely light weight, good fatique and shock resistance, both radiation and moisture resistance, negligible thermal expansion (however, very anisotropic in most cases), low friction constant, etc. In consideration of their still very high cost, RCCs were suggested mainly for nose cones of intercontinental ballistic missiles and wingleading edges of the space shuttles. Some RCCs are also known to be standard components in the brake systems of the top racing cars and look extremely attractive for aviation. RCC samples are rather hard and can be made highly resistant to thermal expansion and temperature gradients if the appropriate construction is achieved. Importantly, to succeed outside the specialty applications of RCC, the production cost, including matrix infiltration and densification, should be drastically reduced. Another important points to address are elastical and thermal isotropies.

There are several common architectures of the carbon – reinforced carbon composite materials. The most spread one is woven graphite fabric laid-up as laminates in two-dimensions. More-dimensional woven structures are of interest but much more expensive than fabric weaves. Other possible architectures include multifilament carbon threads and are referred as tows, chopped discontinuous tows or fabrics, felts, etc. In the present paper, we consider the substitute of the general carbon reinforcement fibres by the ideal-geometry single-walled CNTs to be impregnated into graphite matrix. Molecular dynamics computer simulation method with phenomenological potentials was applied to study the RCC systems at the basic level of structure. The analyzed structures of RCCs were obtained by annealing the pristine graphite and CNTs at high temperatures and pressures. A few molecular dynamics (MD) simulations were conducted employing both van der Waals and intramolecular carbon-carbon interaction potentials. Two thin armchair single-walled CNTs were considered as promising reinforcing agents, CNT (6,6) and CNT (3,3). Structural, mechanical and thermal properties of the resulting RCCs were analyzed in terms of partial atomic density distributions, Young modulus, Poisson's ratio and linear thermal expansion coefficient. We conclude that carbon nanotubes can be found of significant usefulness for RCCs since they allow to obtain extremely light and moderately strong composites with good thermal resistivity.

**Computational Setup**

**Force fields.** The exact composition of all simulated systems is presented in Table 1. Each molecular dynamics system contains a graphite matrix and a certain number of carbon nanotubes, CNT (6,6) or CNT (3,3), of different length. Classical MD simulations with full periodic boundary conditions were performed in order to simulate continuous composite material structures. Gromacs MD engine [28] was used to evaluate intra- and interatomic interactions as well as to integrate motion equations. Further, a few home-made utilities were applied to calculate mechanical properties of the graphite matrix and carbon nanotube – reinforced carbon composites. The interlayer van der Waals interactions in graphite as well as all possible intermolecular interactions within a cutoff (1.2 nm) in the MD systems were simulated using Lennard-Jones (12, 6) carbon-carbon potential. The potential of the same functional form was applied to account for the interactions among carbon atoms which belong to the same structure (nanotube or graphite), but are divided by more than three bonds. The shifted force method with a switch region between 1.1 and 1.2 nm was responsible for the continuity of the pairwise energy function. The natural flexibility of the simulated structures was achieved by means of the terms describing non-harmonic Morse bond, harmonic cosine of the bending angle and two-fold torsion potential. The corresponding force field parameters were successfully used to represent CNTs in the recent publication [29]. Since only chemically identical, non-polarizable carbon atoms are present in all MD systems, no combination rules for Lennard-Jones parameters should be applied. In addition, no electrostatics treatment method is necessary.

**Simulation details**. A graphite matrix was represented using ten initially parallel graphene sheets, ~3.5 nm x 3.0 nm each (4350 atoms). The certain numbers of the nanotubes were randomly impregnated between these sheets to obtain desired compositions of the RCC composites (Table 1). The single-walled armchair CNT (6,6) and CNT (3,3) with the lengths of about 2.43 nm and 0.94 nm, respectively, were considered consequently as reinforcing agents. The lengths of these CNTs were selected in such a way to compare the effects of longer and shorter tubes. The lengths are also smaller than any basic plane dimension of the graphite matrix to avoid calculation of just single molecule mechanical properties. The contents of both CNTs were equal to 5 and 10 wt. %, whereas the same graphite matrix was used in all simulations for clarity. Since the mechanical properties in axial direction of the single nanotube do not explicitly depend on its length [8] (provided that geometry of CNT is perfect), the current simulations setup is quite realistic in spite of the sub-nanometer sizes of simulated objects.

Table 1. The composition of the simulated systems

| N | Material | $N_{sheets}$ | $N_{CNTs}$ | $d_{CNT}$ | $L_{CNT}$ | $N_{atoms}$ |
|---|---|---|---|---|---|---|
| I | 10% CNT (3,3) - RCC | | 13 | 0.41 | 0.94 | 4818 |
| II | 5% CNT (3,3) - RCC | 10 | 6 | | | 4566 |
| III | 10% CNT (6,6) - RCC | | 2 | 0.82 | 2.43 | 4782 |
| IV | 5% CNT (6,6) - RCC | | 1 | | | 4566 |

Constant temperature and constant pressure (NPT) ensemble has been used in these studies. The thermostat proposed by Bussi et al. [30] (often mentioned as V-rescale) with a time constant of 0.1 ps and Parrinello-Rahman barostat [31] with a time constant of 1.0 ps were applied simultaneously in order to maintain the constant temperature and pressure values, respectively. Importantly, the pressure coupling was applied anisotropically to be consistent with an overall geometry of the MD systems. To simulate a moulding stage, the initial structures of the RCCs (combined graphite and nanotubes) were subjected under pressure of 1 000 bar and gradually heated from 300 to 4 000 K during 50 000 ps with a time-step of 0.002 ps in conjunction with a leap-frog integration algorithm. Further, the same procedure was applied for composite cooling. During the moulding stage, the bond lengths were maintained constant using the LINCS algorithm. In turn, the mechanical properties (strain-stress curves, Young modulus and Poisson's ratio) were calculated at 5 K in order to decrease the role of thermal fluctuations. The boundary atom layers were fixed after the strain is applied, and the time-step was 0.0001 ps for all strained systems, with all bond constraints removed.

**Data analysis**. The structure of RCC composites was characterized in terms of partial densities of the nanotubes along each axis within a graphite matrix (Figure 2).

Young modulus, also known as the tensile modulus (Figure 3, Table 2), $E$, was calculated as the second derivative of the total internal energy, $E_{tot}$, with respect to the applied strain, $\varepsilon$, provided that strain is low,

$$E = -\frac{1}{V}\left(\frac{d^2 E_{tot}}{d\varepsilon^2}\right)_{\varepsilon \to 0}, \qquad (1)$$

where $V$ is an equilibrium volume of the system under strain.

In detail, the simulations of tensile modulus were performed by applying a certain tension along the chosen direction and fixing the corresponding edges of the periodic sample, while the structure is allowed to relax with respect to its new geometry. The equilibrium energies were used to estimate the corresponding Young modulus using (eq. 1). The size of the edge layer was selected to be 0.5 nm, somewhat exceeding single carbon atom van der Waals diameter. Neither inter-, nor intramolecular interaction energies of these fixed edges were accounted during evaluation of Young modulus and Poisson's ratio.

Poisson's ratio (Figure 4, Table 3), $\nu$, is a ratio, when a sample object is stretched, of the contraction or transverse strain (perpendicular to the applied load), $\varepsilon_{trans}$, to the extension or axial strain (in the direction of the applied load), $\varepsilon_{axial}$,

$$\nu = \left(-\frac{\varepsilon_{trans}}{\varepsilon_{axial}}\right)_{\varepsilon \to 0}. \qquad (2)$$

Poisson's ratio is a standard measure of the Poisson's effect which is caused by slight movements between particles and stretching of molecular bonds within the material lattice to accommodate the stress. Thus, it can be reliably derived from atomistic simulations using the constant pressure ensemble. The Poisson's ratio of any stable, isotropic, linear elastic material cannot be neither less than −1.0 nor greater than 0.5, while most material (steels, rigid polymers) exhibit the values ranging between 0.0 and 0.5. However, for anisotropic materials this does not apply, and the possible Poisson's ratio has no theoretical limits.

The coefficient of thermal expansion (Figure 4, Table 4), $\alpha$, is another important property of composite material, especially if it is intended to be used over large temperature range. Usually, the materials expand as temperature grows since it leads to greater thermal vibration of the atoms, and hence to increase in the average separation distance of the adjacent particles. The volumetric coefficient of thermal expansion accounts to the change of volume, V, with a temperature, T, change,

$$\alpha_V = \frac{1}{V}\frac{dV}{dT}, \qquad (3)$$

In turn, for anisotropic materials, linear coefficients of thermal expansion, $\alpha_x$, $\alpha_y$, $\alpha_z$ give a better description of the material. In order to calculate $\alpha_x$, $\alpha_y$, $\alpha_z$, the RCC samples were gradually heated from 5 to 1 000 K during 50 000 ps, and the lengths of the cell edges were monitored. The corresponding coefficients have been derived from the slopes of the thermal expansion plots at 200 and 1000 K (Figure 4, Table 4).

**Results and discussion**

**Density and structure**. Generally, the density of graphite varies very significantly, from 0.2 g/cm$^3$ for a high-porous case to 2.2 g/cm$^3$ for pyrolitic graphite. The simulated sample, built up using ten uniform parallel graphene flakes with pure van der Walls interactions between them, yields a density of ~2.2 g/cm$^3$ at normal conditions. Graphite of such density is usually obtained *via* thermomechanical treatment of the natural species. Note that presently simulated graphite contains no structural defects and its density is significantly higher than conventional graphite matrix should exhibit.

The densities of the single-walled CNTs are generally defined by their effective diameters (curvatures) and lengths. As the nanotube is basically a rolled up graphene sheet, its density is expected between the above-mentioned minimal and maximal values for graphite. The simulated densities of CNT (6,6) and CNT (3,3) bundles are 1.40 and 1.26 g/cm$^3$, respectively. Although CNT (3,3) has a very small inner cavity, the density of this sample is slightly lower than of the CNT (6,6) bunch. Such curious fact can be attributed to the nanotube length since CNT (6,6) contains significantly more chemically bonded carbon atoms (216 instead of 36) with a smaller interatomic distance between them.

The densities of all RCCs (Table 2) are similar, with larger CNT loadings leading to an insignificant density decrease. The density of RCC is determined by the general architecture of composite material and should not directly depend on the density of the CNT bunches. The extreme lightness (1.4–1.5 g/cm$^3$) of the considered RCC samples, which are 4-5 times lighter than steel, greatly favor their use in aviation and space shuttles. Another important factors, needed for these specialty applications, are reasonable stiffness and low thermal expansion coefficients in a wide temperature range.

Table 2. Densities, $d_{5K}$, $d_{200K}$, $d_{1000K}$, of the RCC samples derived from MD simulations at 5, 200, 1 000 K, respectively, and Young moduli, $E_x$, $E_y$, $E_z$, along Cartesian directions at 5 K and 1 bar

| N | $d_{5K}$ | $d_{200K}$ | $d_{1000K}$ | $E_x$ | $E_y$ | $E_z$ |
|---|---|---|---|---|---|---|
| I | 1.51 | 1.50 | 1.48 | 0.74 | 0.13 | 0.94 |
| II | 1.48 | 1.48 | 1.46 | 0.51 | 0.15 | 0.64 |
| III | 1.42 | 1.41 | 1.39 | 0.45 | 0.16 | 0.81 |
| IV | 1.49 | 1.48 | 1.46 | 0.69 | 0.41 | 0.04 |

The partial atomic densities of the carbon nanotubes along each Cartesian direction, x, y and z, are depicted in Figure 1. At nanoscopic level, the absolute instantaneous values vary appreciably, between 0 and 900 kg/m$^3$. It means that in some local regions of the RCC sample the density of nanotubes is more than 50%, whereas the average density of CNTs is 5-10% resulting in poor dispersion. In the case of smaller nanotube, CNT (3,3), the obtained material is more isotropic (Figure 1 (a, b)), whereas in the case of CNT (6,6) some directions (Figure 1 (c, d)) exhibit very sharp transitions of the local atomic densities. From this point, it is questionable if longer CNTs (note, CNT (6,6) is twice shorter than the simulated graphene layer) are in preference to shorter ones in order to obtain isotropic samples. Although theoretically long nanotubes can be more efficient as reinforcing agents due to a larger density of chemical bonds (and hence, potential energy), special attention should be paid to distribute them uniformly over the entire graphite matrix. Herein, specific production techniques are urgently needed to improve dispersion of the reinforcing agent. Chemical functionalization of both nanotubes and matrix can also be of interest to increase the affinity between the composite building elements.

**Mechanical properties**. The full relaxation of the RCC structure, subjected to deformation, occurs within 5 ps. The equilibrium total energies at different strains were used to derive Young modulus. The strain-stress curves for x, y and z directions separately derived for system I (Table 1) are shown in Figure 3. This dependence is linear when strain is less than 2% and can be well described by means of the second-order polynomial when strain is beyond the elastic limit. In order to estimate tensile modulus (Table 2) *via* (eq. 1), we used the strains of 0.8, 0.9, 1.0, 1.1 and 1.2% along each direction to be sure that Hooke's law holds. The larger strain values will not be further considered in this paper, since more sophisticated many-body interaction potentials must be applied to simulate this region of strains adequately.

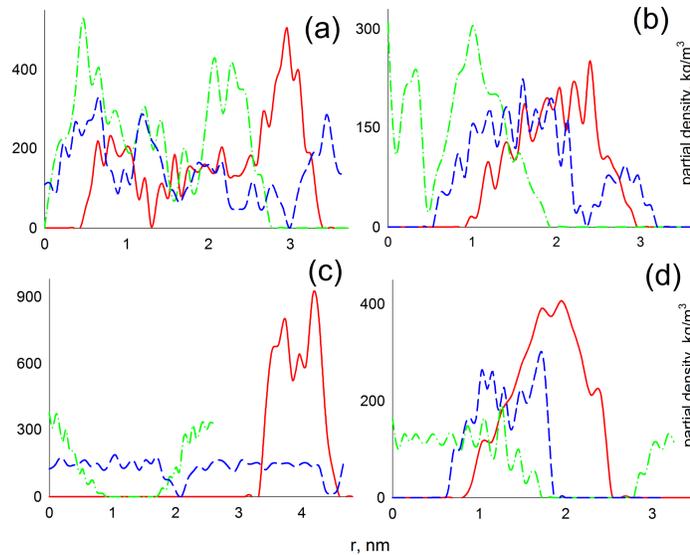

Figure 1. (Color online) The partial densities of carbon nanotubes along Cartesian directions, z (solid red), y (dash-dotted green) and x (dashed blue) in the simulated RCC samples, I (a), II (b), III (c) and IV (d).

In general, the bigger fraction of the CNTs in the RCC sample leads to the Young modulus increase (Figure 2). This observation is true both for CNTs (3,3) and CNTs (6,6) impregnated into the same matrix containing ten graphene layers. Interestingly, the results exhibited by shorter nanotubes (systems I-II) are somewhat better than those by longer ones (systems III-IV).

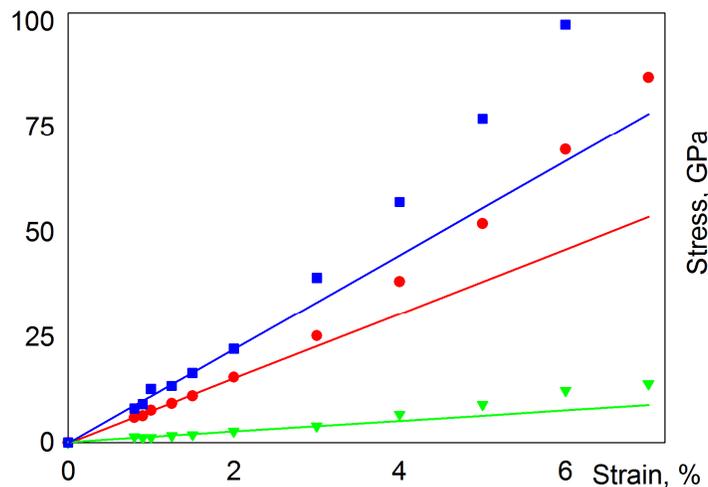

Figure 2. (Color online) The Young modulus along Cartesian directions, x (red circles), y (green triangles) and z (blue squares) of the simulated RCC samples.

It may contradict the intuition since longer tubes contain more chemically bonded atoms resulting in a higher potential energy densities, and stiffness. Importantly, the present outcome suggests that fine dispersion (Figure 2) of the CNTs within a matrix appears to be more important than the particular geometrical parameters of the nanotubes. The addition of the nanotubes allows to achieve up to 8 times increase of the tensile modulus along the graphite interlayer direction. However, the correlation of this value with CNT loading is not clear. The observed Young modulus (0.13-0.16 TPa) in the interlayer direction are still worse than most steels exhibit (~0.3 TPa). Nevertheless, we believe this drawback is completely compensated by the RCC's very low weights (~1.4-1.5 g/cm$^3$). In the directions of basic graphite planes, the tensile modulus ranges between 0.41-0.95 TPa (Table 2) and is 7-60% smaller than of the single graphene layer (1.025 TPa [32]). Note, in practical applications the corresponding modulus should probably be even lower because of the structure non-ideality.

Unfortunately, the calculated Young modulus is rather anisotropic for all samples. In fact, the observed anisotropy originates in the highly anisotropic properties of the graphite matrix which contains no chemical bonds along its interlayer direction. Interestingly, the fraction of the nanotubes does not significantly influence anisotropy. Such very important practical issue can probably be addressed with more sophisticated production techniques than simple moulding stage as applied in the reported study.

Poisson's ratio (eq. 2) along with Young modulus is often used to describe the elastic properties of the material. The Poisson's ratios for each direction, $v_x$, $v_y$, $v_z$, calculated at different strains are depicted in Figure 3 (system I), and the corresponding values for all samples are listed in Table 3. When low deformations occur, all Poisson's ratios are positive, although at bigger deformations the Poisson's ratios in basic planes decrease sharply and even become negative at strains bigger than 2%. For interlayer direction, this tendency is rather weakly pronounced (Figure 3).

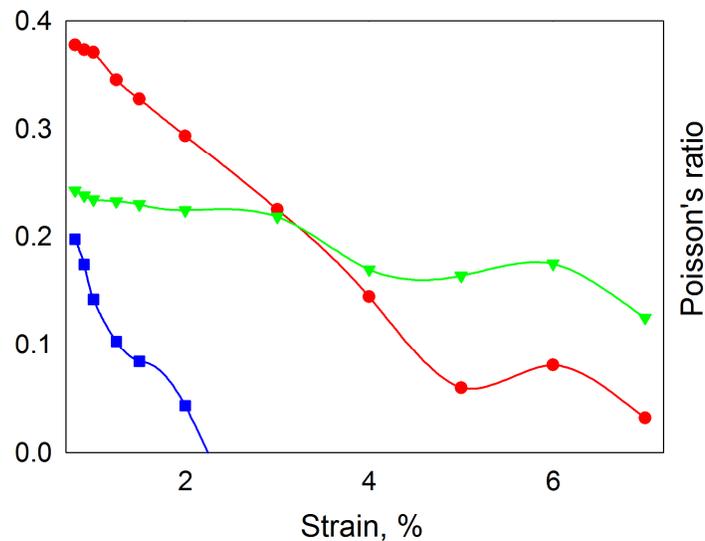

Figure 3. (Color online) The Poisson's ratio along Cartesian directions, x (red circles), y (green triangles) and z (blue squares) of the simulated RCC samples.

As well, as the Young modulus, the Poisson's ratio, $v$ (calculated at $\varepsilon \rightarrow 0$) of the different RCC samples are very anisotropic. Although all values of $v$ are positive, they can differ more than three times for perpendicular tensions. For bigger CNT loadings, $v_x$, $v_y$, $v_z$ are somewhat smaller and close to one another. These values satisfactorily correlate with the properties of most steels ($v \approx 0.30$) and graphite ($v \approx 0.20-0.27$). For 5% fraction, the scattering of the Poisson's ratios can be probably attributed to insufficient simulation sampling due to very small number of carbon nanotubes (6 and 1 for systems II and IV, respectively). This can also result in an increased anisotropy of the RCC samples on the nanometer scale, that is not noticeable in the macroscopic scales.

Table 3. Poisson's ratio of the RCC samples derived from MD simulations at 5 K and 1 bar.

| N   | $\nu_x$ | $\nu_y$ | $\nu_z$ |
|-----|---------|---------|---------|
| I   | 0.39    | 0.23    | 0.21    |
| II  | 0.61    | 0.30    | 0.52    |
| III | 0.24    | 0.32    | 0.17    |
| IV  | 0.48    | 0.38    | 0.33    |

The joint analysis of Young modulus (Table 2) and Poisson's ratio (Table 3) of the simulated RCC samples (Table 1) show that the mechanical properties of the constructed carbon nanotube – reinforced carbon composites are somewhat worse than of steels (due to the graphite interlayer direction). Although for many applications this can be thoroughly compensated by the RCC's extremely low weights (1.4-1.5 g/cm$^3$).

**Thermal expansion**. The coefficient of thermal expansion, $\alpha$, of pure graphite matrix changes with temperature and is known to be highly anisotropic. In basic planes, $\alpha = -1.2 \times 10^{-6} K^{-1}$ at T < 200 K, $\alpha \approx 0$ at 700 K < T < 1000 K and $\alpha = 0.7 \times 10^{-6} K^{-1}$ at T > 1000 K. In perpendicular plane, $\alpha$ is more than 20 times larger and weakly dependent on temperature since only weak van der Waals interactions are present in this direction. This anisotropy of thermal expansion is transferred to the composite materials based on graphite matrix. Thus, three linear coefficients are of more interest than volumetric one (Table 4).

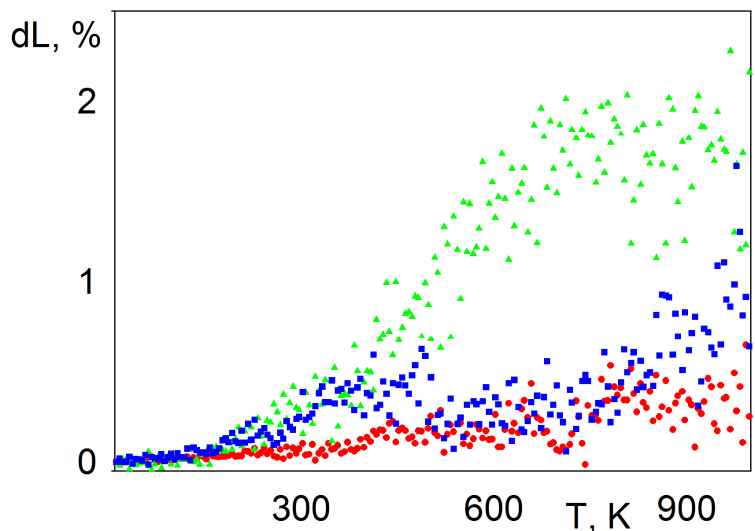

Figure 4. (Color online) The temperature dependences of the RCC's relative linear dimensions, x (red circles), y (green triangles) and z (blue squares) of the simulated RCC sample (system I).

The relative variations of the RCC volumes at temperature increase from 5 to 1 000 K are shown in Figure 4. The thermal expansion of the RCCs is not constant as temperature ranges from 5 to 1000 K and is very close to zero at temperatures below 200 K. As well as in the graphite matrix itself, the largest $\alpha$ is found along the direction perpendicular to basic planes where it is positive at all temperatures. In a whole, the linear coefficients of thermal expansion well correlate with Young modulus values in the same directions being in inverse ratio to them. Although $\alpha$ of RCC composites (Table 4) is generally higher then those of graphite, the values of 10$^{-6}$-10$^{-5}$ K$^{-1}$ can still be considered satisfactory for RCCs to be used at large temperature gradients. Since the biggest $\alpha$ is observed in the case of interlayer direction, this property will certainly be improved if one succeeds to decrease anisotropy of RCC.

Table 4. The linear coefficients of thermal expansion ($10^5$ K$^{-1}$) of the RCC samples derived from MD simulations at 200 and 1 000 K

| N | $\alpha_x$ | $\alpha_y$ | $\alpha_z$ | $\alpha_x$ | $\alpha_y$ | $\alpha_z$ |
|---|---|---|---|---|---|---|
| | | 200 K | | | 1 000 K | |
| I | 0.4 | 1.8 | 0.3 | 0.5 | 2.3 | 0.7 |
| II | -0.3 | 1.9 | -0.08 | 0.8 | 2.1 | 0.8 |
| III | 0.6 | 5.5 | 0.4 | 0.7 | 5.0 | 0.4 |
| IV | -0.6 | 1.3 | 4.0 | 0.3 | 1.1 | 6.3 |

**Conclusions**

To recapitulate, we have performed molecular dynamics simulations of the carbon nanotube – carbon reinforced composites. Partial CNT densities within graphite matrix, Young modulus, Poisson's ratio and linear thermal expansion coefficients were calculated in order to characterize the obtained samples. The addition of CNTs allows to increase Young modulus in the interlayer direction of graphite matrix up to 8 times and decrease the sample weight by 30-40 %. The linear thermal expansion coefficients are of the order of $10^{-6}$-$10^{-5}$ K$^{-1}$ that favors application of RCC in aviation and space vehicles. The dispersion of the CNTs within matrix is shown to play a crucial role in the composite mechanical and thermal properties. The current drawback is a significant anisotropy of the analyzed RCC samples. Further efforts are urgently needed to decrease it.

**Acknowledgment**

This is a great pleasure to thank Julia Nazarenko (National Aerospace University, Kharkiv, Ukraine) for the inspiration to conduct this research.

# Literature

1. G. Hummer, J.C. Rasaiah, J.P. Noworyta, Nature. 414 (2001) 188.
2. S. Singh, J. Houston, F. van Swol, C.J. Brinker, Nature. 442 (2006) 526.
3. M.C. Gordillo, J. Marti, Chem. Phys. Lett. 329 (2000) 341.
4. M.C. Gordillo, J. Boronat, J. Casulleras, Phys. Rev. B. 61 (2000) R878.
5. J. Marti, M.C. Gordillo, J. Chem. Phys. 119 (2003) 12540.
6. P.M. Ajayan, J.M. Tour, Nature. 447 (2007) 1066.
7. P. Calvert, Nature. 399 (1999) 210.
8. M.M. Shokrieh, R. Rafiee, Mechanics of Composite Materials. 46 (2010) 155.
9. A. Ghosh, S. Ghosh, S. Das, P.K. Das, D.D. Majumder, R. Banerjee, Chem. Phys. Lett. 496 (2010) 321.
10. K.P. Ryan, S.M. Lipson, A. Drury, M. Cadek, M. Ruether, S.M. O'Flaherty, V. Barron, B. McCarthy, H.J. Byrne, W.J. Blau, J.N. Coleman, Chem. Phys. Lett. 391 (2004) 329.
11. M. Dutta, D. Basak, Chem. Phys. Lett. 480 (2009) 253.
12. C. Pirlot, Z. Mekhalif, A. Fonseca, J.B. Nagy, G. Demortier, J. Delhalle, Chem. Phys. Lett. 2003. 372(3-4): p. 595-602.
13. X. Sun, R. Li, D. Villers, J.P. Dodelet, S. Desilets, Chem. Phys. Lett. 379 (2003) 99.
14. D. Qian, E.C. Dickey, Journal of Microscopy-Oxford. 204 (2001) 39.
15. A. Garg, S.B. Sinnott, Chem. Phys. Lett. 295 (1998) 273.
16. X. Li, C.Y. Li, X.M. Li, H.W. Zhu, J.Q. Wei, K.L. Wang, D.H. Wu, Chem. Phys. Lett. 481 (2009) 224.
17. Y.H. Li, J.Q. Wei, X.F. Zhang, C.L. Xu, D.H. Wu, L. Lu, B.Q. Wei, Chem. Phys. Lett., 365 (2002) 95.
18. S.L. Mielke, S. Zhang, R. Khare, R.S. Ruoff, T. Belytschko, G.C. Schatz, Chem. Phys. Lett. 446 (2007) 128.
19. G. Zhou, W.H. Duan B.L. Gu, Chem. Phys. Lett. 333 (2001) 344.
20. K. Koziol, J. Vilatela, A. Moisala, M. Motta, P. Cunniff, M. Sennett, A. Windle, Science, 318 (2007) 1892.
21. L. Sun, F. Banhart, A.V. Krasheninnikov, J.A. Rodriguez-Manzo, M. Terrones, P.M. Ajayan, Science. 312 (2006) 1199.
22. P.M. Ajayan, O. Stephan, C. Colliex, D. Trauth, Science. 265 (1994) 1212.
23. C.W. Fan, Y.Y. Liu, C. Hwu, Applied Physics a-Materials Science & Processing, 95 (2009) 819.
24. G. Soldano, M.M. Mariscal, Nanotechnology. 20 (2009) 165705.
25. M. Motta, Y.L. Li, I. Kinloch, A. Windle, Nano Lett. 5 (2005) 1529.
26. C.H. Ke, M. Zheng, G.W. Zhou, W.L. Cui, N. Pugno, R.N. Miles, Small. 6 (2010) 438.
27. J. Yang, L.Q. Zhang, J.H. Shi, Y.N. Quan, L.L. Wang, M. Tian, J. Appl. Polymer Science, 116 (2010) 2706.
28. B. Hess, C. Kutzner, D. van der Spoel, E. Lindahl, J. Chem. Theory and Comput. 4 (2008) 435.
29. J. H. Walther, R. Jaffe, T. Halicioglu, P. Koumoutsakos, J. Phys. Chem. B. 105 (2001) 9980.
30. G. Bussi, D. Donadio, M. Parrinello, J. Chem. Phys. 126 (2007) 014101.
31. M. Parrinello, A. Rahman, J. Appl. Phys. 52 (1981) 7182.
32. B. WenXing, Z. ChangChun, C. WanZhao, Physica B. 352 (2004) 156.